\documentclass[aps,prc,reprint]{revtex4-1}

\usepackage{amsmath, amssymb}
\usepackage[pdftex]{graphicx}

\newcommand{\be}{\begin{equation}}
\newcommand{\ee}{\end{equation}}
\newcommand{\ba}{\begin{eqnarray}}
\newcommand{\ea}{\end{eqnarray}}
\newcommand{\bd}{\begin{displaymath}}
\newcommand{\ed}{\end{displaymath}}

\def\oneth{{\textstyle{\frac{1}{3}}}}
\def\twoth{{\textstyle{\frac{2}{3}}}}

\begin{document}

\title{Departure from Equilibrium of the Quasi-Particle Distribution Functions in High Energy Nuclear Collisions}
\author{P. Chakraborty}
\affiliation{Department of Physics, Basirhat College, Basirhat, West Bengal, India}
\author{J. I. Kapusta}
\affiliation{School of Physics and Astronomy, University of Minnesota,
Minneapolis, Minnesota 55455, USA}



\begin{abstract}
In simulations of high energy heavy ion collisions that employ viscous hydrodynamics, single particle distributions are distorted from their thermal equilibrium form due to gradients in the flow velocity.  These are closely related to the formulas for the shear and bulk viscosities in the quasi-particle approximation.  Distorted single particle distributions are now commonly used to calculate the emission of photons and dilepton pairs, and in the late stage to calculate the conversion of a continuous fluid to individual particles.  We show how distortions of the single particle distribution functions due to both shear and bulk viscous effects can be done rigorously in the quasi-particle approximation and illustrate it with the linear $\sigma$ model at finite temperature.
\end{abstract}

\maketitle

\section{Introduction}

A standard theoretical calculation of a high energy heavy ion collision at the BNL Relativistic Heavy Ion Collider (RHIC) or CERN Large Hadron Collider (LHC) proceeds as follows.  A model is chosen for a description of the state of matter soon after impact, typically after a few tenths to one fm/c in proper time.  This provides the initial condition for the subsequent hydrodynamic evolution.  When the system becomes sufficiently dilute, typically when the local temperature falls to somewhere in the range $120 < T < 150$ MeV, the continuous fluid is converted into individual hadrons using the so-called Cooper-Frye formula.  Sometimes this is the final predicted spectrum of hadrons, and sometimes this is followed by an after-burner which allows the hadrons to scatter until they ultimately fly off to the detectors.  For a good overview of the field see the Proceedings of the Series of Quark Matter Meetings, the most recent ones being \cite{QM2014,QM2015}.

Departures from the local equilibrium distribution $f_a^{\rm eq}$ for particle species $a$ with momentum ${\bf p}$ are expressed in terms of the function $\phi_a$ as 
\be
f_a = f_a^{\rm eq} \left( 1 + \phi_a \right) \,.
\ee
In the local rest frame
\ba
\phi_a &=& -A_a \nabla \cdot \boldsymbol{v} +  C_a p^ip^j \left( \partial_i v_j +  \partial_j v_i + \twoth \delta_{ij} \nabla \cdot \boldsymbol{v} \right) \nonumber \\
&=& -A_a \nabla \cdot \boldsymbol{v} +  2 C_a p^ip^j \left( \partial_i v_j + \oneth \delta_{ij} \nabla \cdot \boldsymbol{v} \right) \, .
\ea
The local fluid velocity $\boldsymbol{v}$ depends on space and time while the two scalar functions $A_a$ and $C_a$ depend on the momentum of the particle.  (Note that here and throughout we assume zero baryon chemical potential.)  The functions $A_a$ are associated with bulk viscosity and the functions $C_a$ are associated with shear viscosity.  The out of equilibrium distribution functions are frequently used nowadays to calculate a number of observables, such as hadron spectra and elliptic flow \cite{Teaney-df} and photon and dilepton production \cite{em-df1}.  The lowest order approximation is to assume that $C_a$ is independent of momentum and particle species \cite{degroot} and was introduced into the field of heavy ion collisions in \cite{Teaney-df}.  For independent particles it results in the expression
\be
C_a = \frac{\eta/s}{2T^3} \,,
\ee
where $\eta$ is the shear viscosity and $s$ is the entropy density.  One also assumes all $A_a = 0$.  An exploratory study to go beyond the assumption that $C_a$ is independent of momentum and particle species was reported in \cite{MolnarWolff}.  The effects of $A_a \neq 0$ on photon production was reported in \cite{em-df2} under the simplifying assumptions of equal relaxation times for all particles.

In this paper we study more accurate expressions for the functions $A_a$ and $C_a$, especially their dependence on momentum and on particle species.  This is particularly important if one wants to learn more about the transport properties of high temperature matter from experimental data as mentioned above.  Our study is limited to quasi-particle models; the linear $\sigma$ model is used to illustrate the magnitude of numerical departures from Eq. (3) and from $A_a = 0$.  Results for resonance gas models, use of the more sophisticated Chapman-Enskog approximation, and inclusion of chemical potentials are postponed to future work.  Section II studies shear effects, Sec. III studies bulk effects, and Conclusions are presented in Sec. IV.  Throughout the paper we use the notation, derivations, and results of \cite{transport1}.  

\section{Shear Viscosity}


\begin{figure*}[t]
\centering
\includegraphics[width=0.48\textwidth]{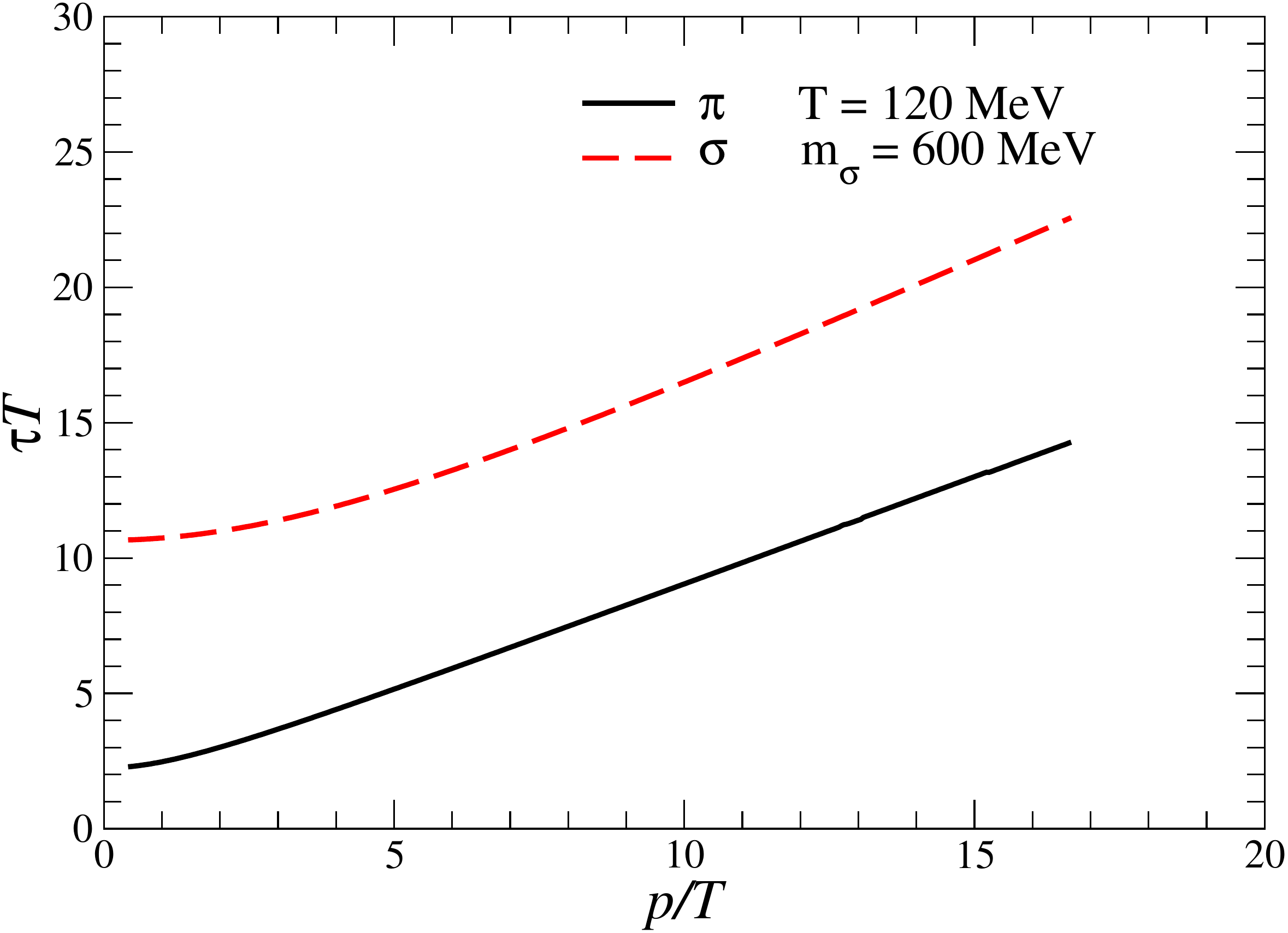}
\includegraphics[width=0.48\textwidth]{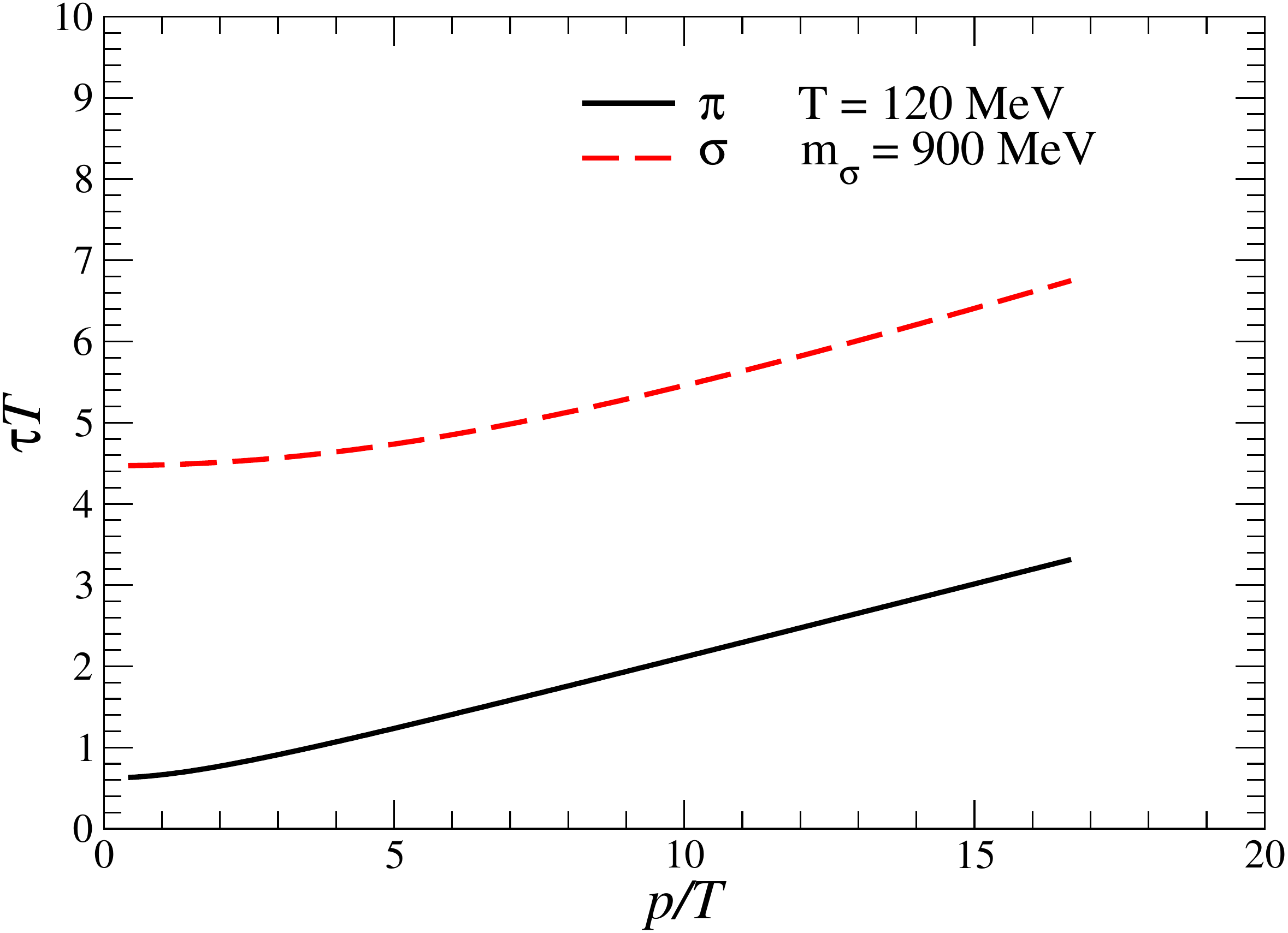}
\includegraphics[width=0.48\textwidth]{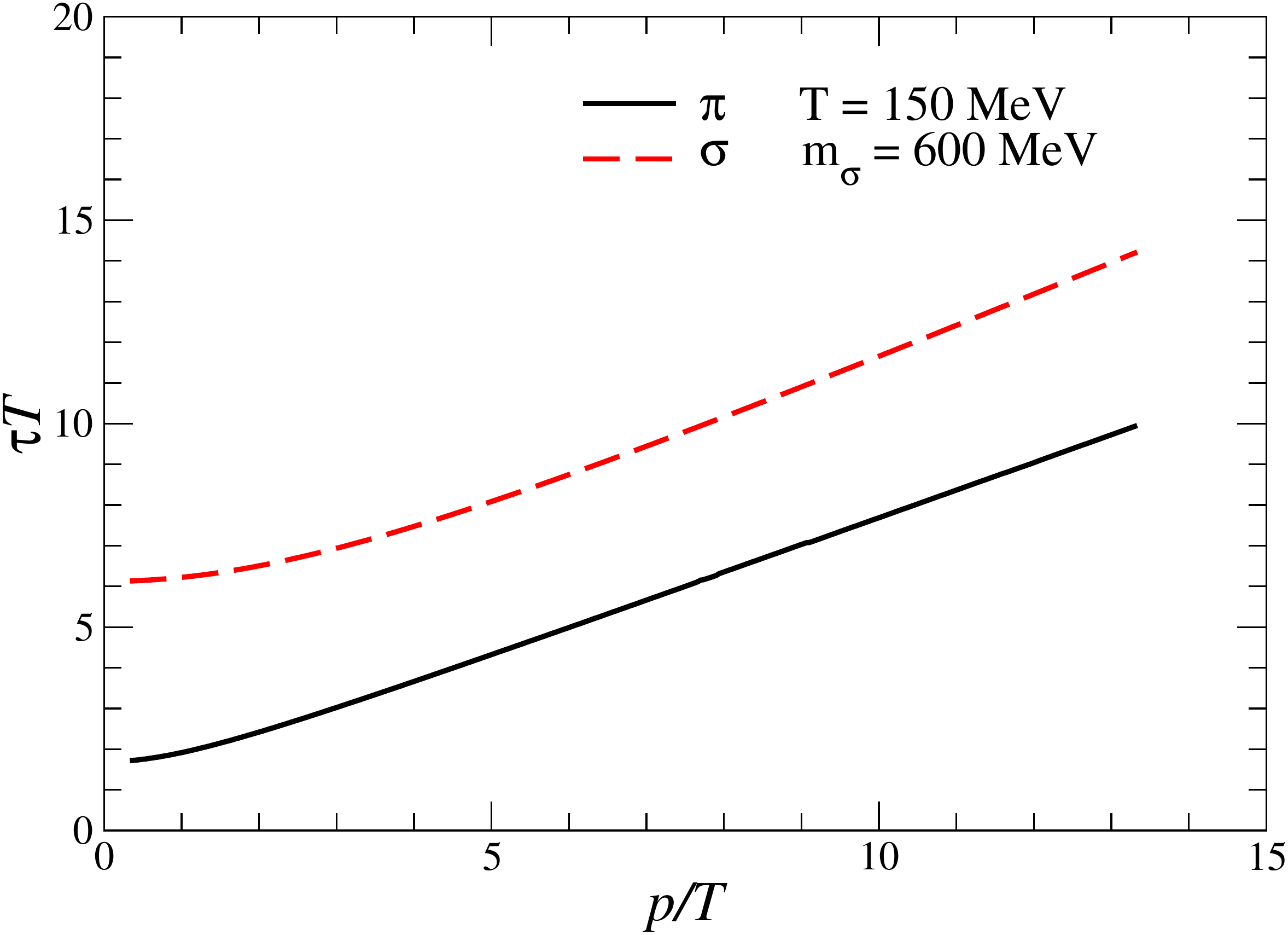}
\includegraphics[width=0.48\textwidth]{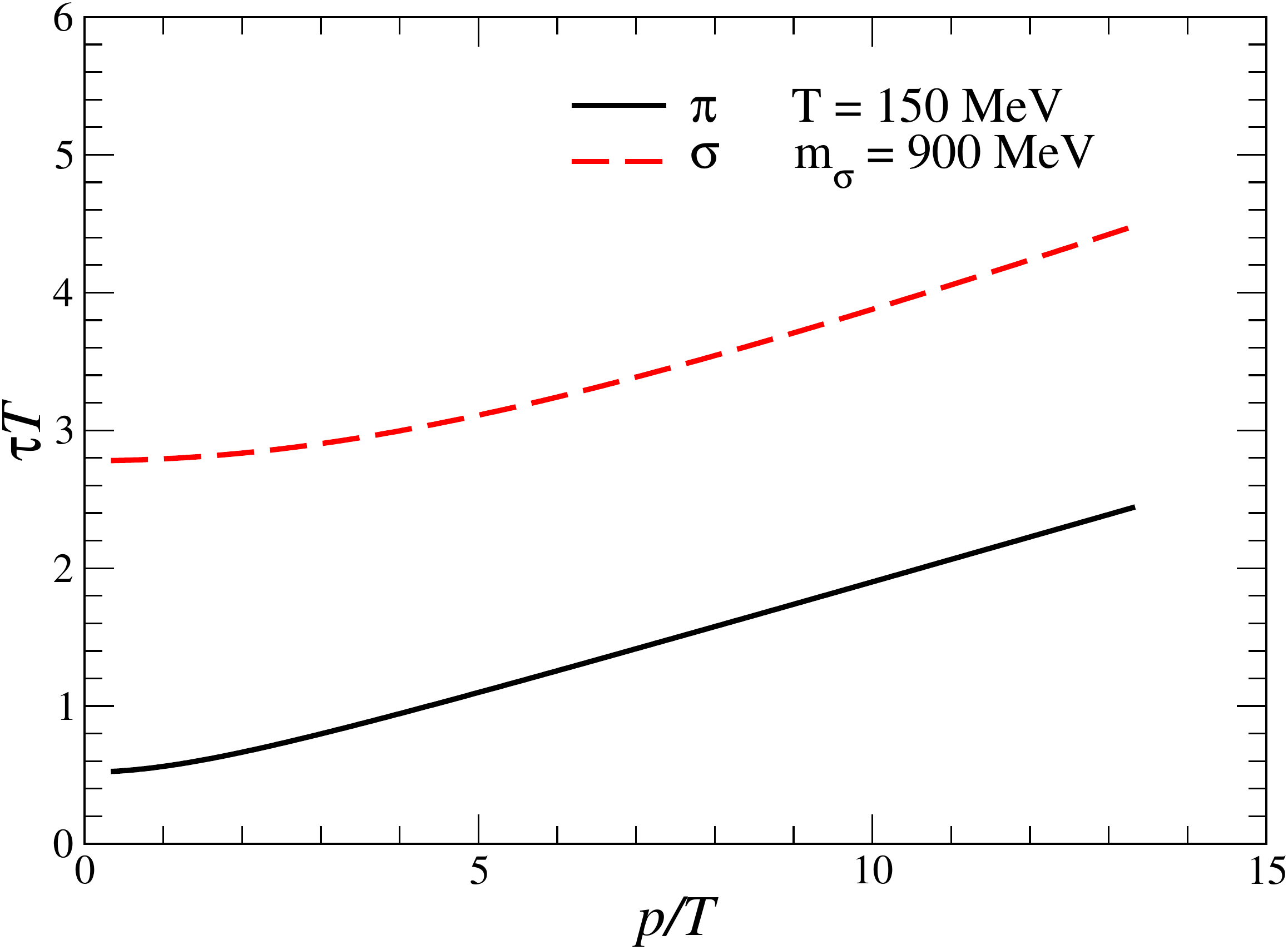}
\caption{Relaxation times $\tau_{\pi}$ and $\tau_{\sigma}$ in the linear $\sigma$ model for the indicated values of the vacuum $\sigma$ mass and temperature.}
\end{figure*}

In quasi-particle models, which may include mean fields affecting the effective, temperature-dependent masses $\bar{m}_a(T)$, the shear viscosity is expressed in terms of the functions $C_a$ as
\be
\eta = \frac{2}{15} \sum_a \int \frac{d^3p}{(2\pi)^3} \frac{|{\bf p}|^4}{E_a} 
f_a^{\rm eq}(E_a/T) C_a(E_a)
\label{gasshear}
\ee
where $E_a = \sqrt{{\bf p}^2 + \bar{m}_a^2(T)}$.   In quasi-particle models there is a mean field but the expression for the entropy density has the same form as for non-interacting particles, with vacuum masses replaced by temperature-dependent masses.  Specifically, the pressure is $P = P_0 - V$ and the energy density is $\epsilon = \epsilon_0 + V$, where the subscript 0 refers to free particle formulas with vacuum masses replaced by temperature-dependent masses, and $V$ is the potential energy density.  Thus $T s = P+\epsilon = P_0 +\epsilon_0 = T s_0$.  If one assumes that all $C_a \equiv C$ are independent of species and momentum, then it can be factored out of the integral with the result that
\be
\eta = 2 T^3 s C \,,
\ee
which is the same as Eq. (3).

In the momentum-dependent relaxation time approximation, we have
\be
C_a = \frac{\tau_a(E_a)}{2 T E_a} \,,
\ee
where $\tau_a(E_a)$ is the relaxation time for particles of species $a$.  From Eqs. (4) and (6) one sees that the particle species $a$ will make a significant contribution to the shear viscosity if it has a large relaxation time and has a small mass.  Physically it means that such a particle can transport energy and momentum over a large distance and that they are relatively abundant.

As an illustration we use the linear $\sigma$ model as in \cite{transport1}.  Figure 1 shows the relaxation times for the pion and $\sigma$ meson for $m_{\sigma} = 600$ and 900 MeV, and for $T = 120$ and 150 MeV, as functions of the momentum.  The pion has a shorter relaxation time which is not surprising.  The relaxation times increase with momentum which is not surprising either.

\begin{figure*}[t]
\centering
\includegraphics[width=0.48\textwidth]{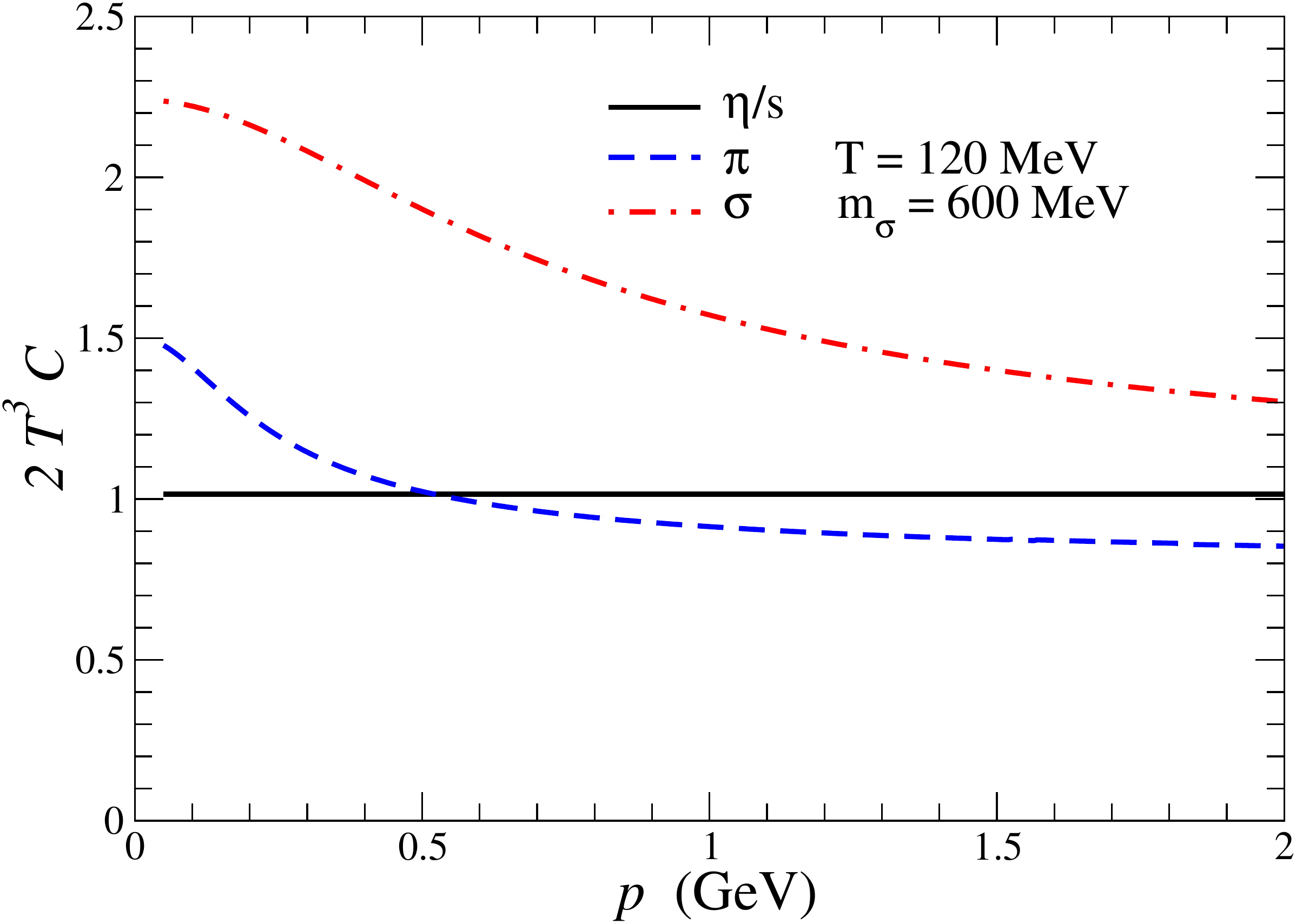}
\includegraphics[width=0.48\textwidth]{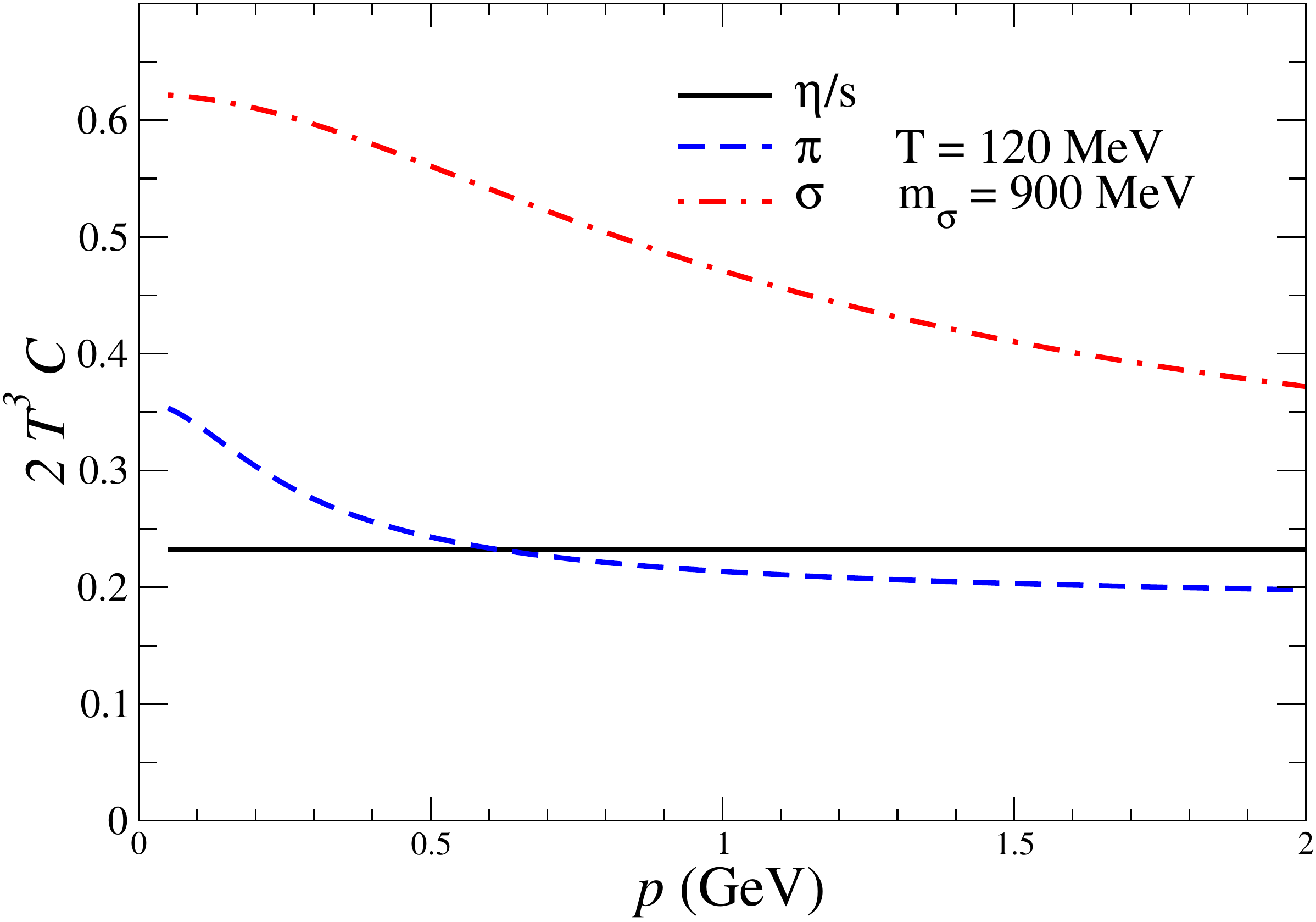}
\includegraphics[width=0.48\textwidth]{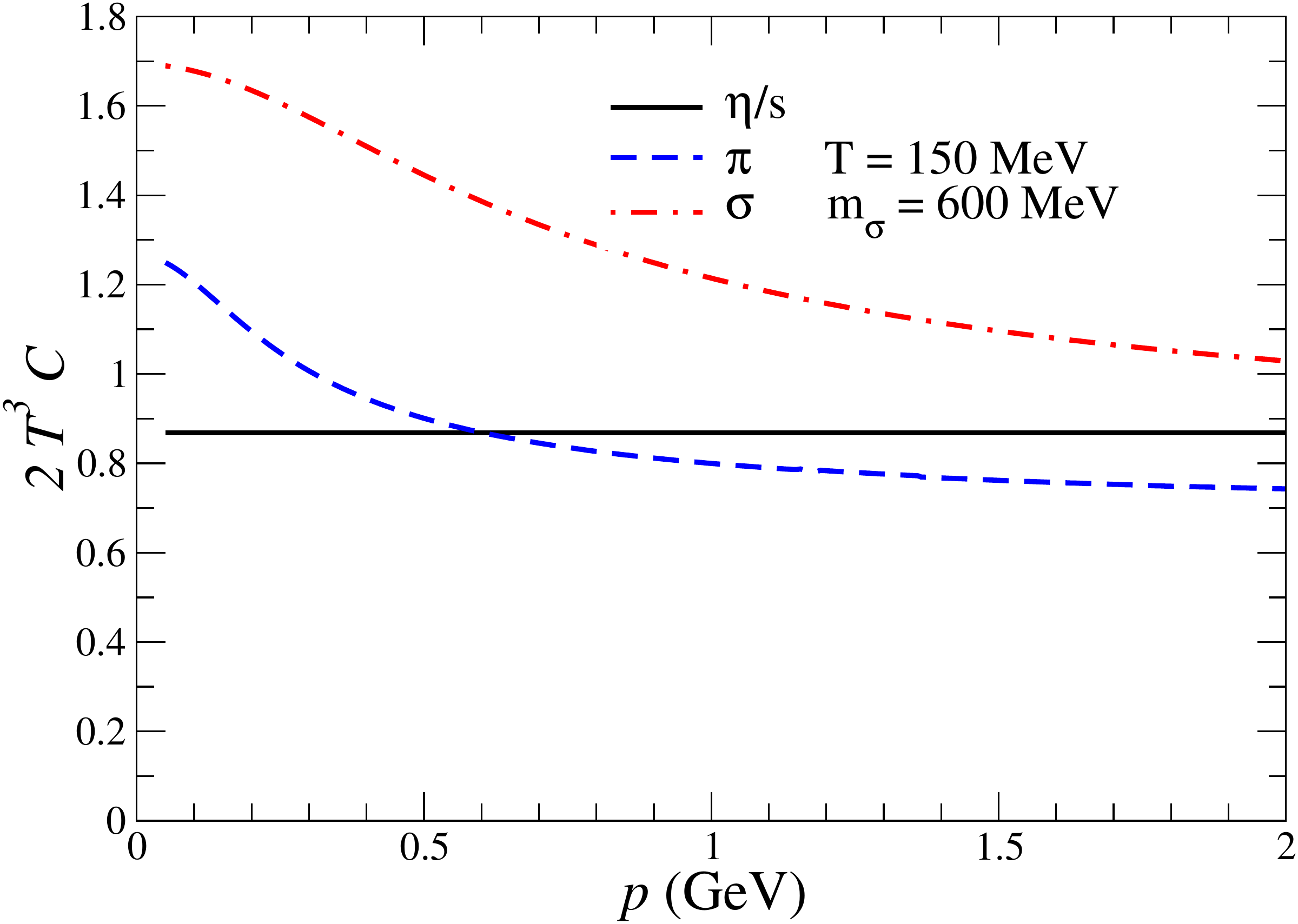}
\includegraphics[width=0.48\textwidth]{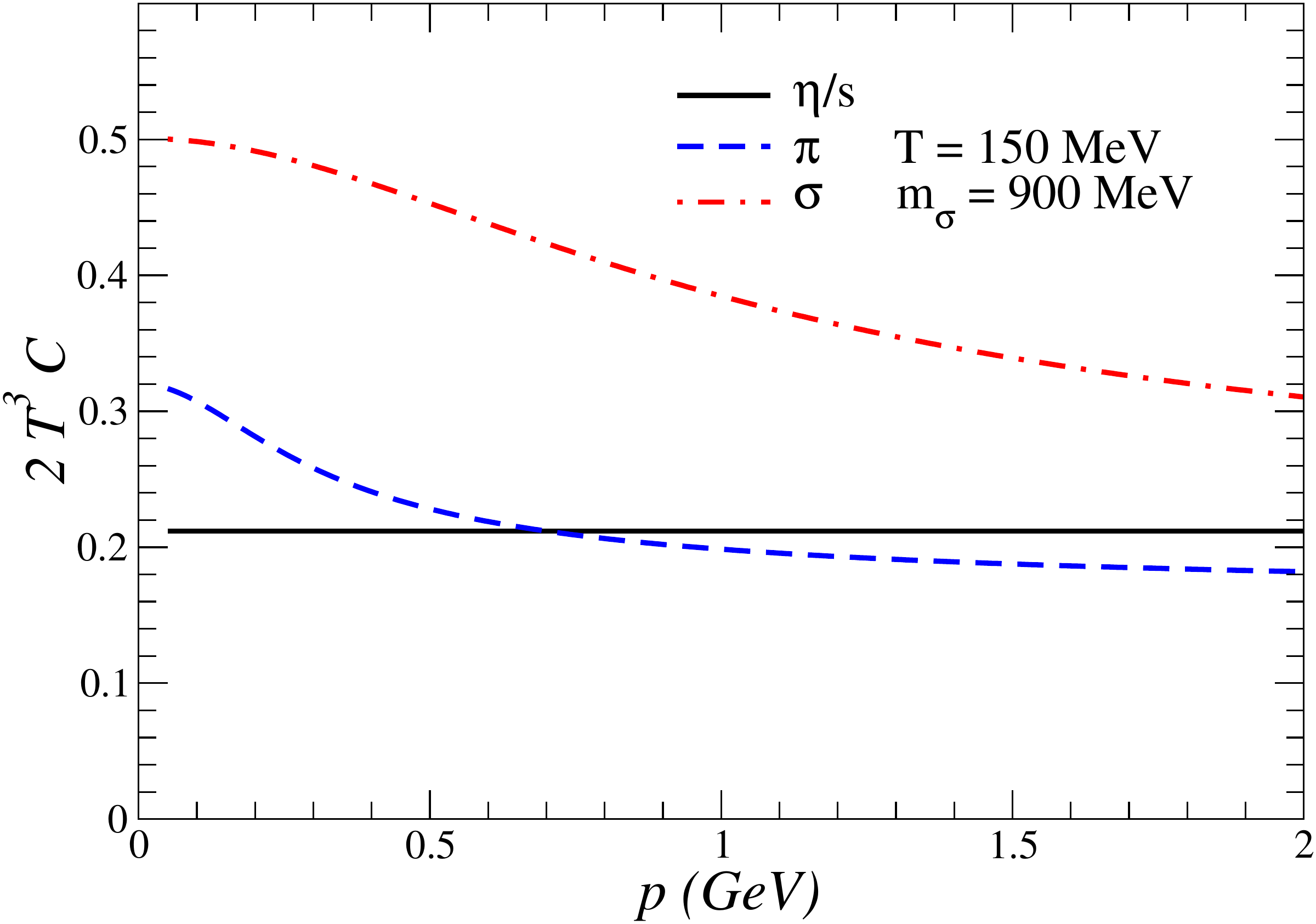}
\caption{The shear functions $C_{\pi}$ and $C_{\sigma}$ in the linear $\sigma$ model for the indicated values of the vacuum $\sigma$ mass and temperature.}
\label{Fig2}
\end{figure*}

Figure \ref{Fig2} shows the resulting functions $C_{\pi}$ and $C_{\sigma}$ as well as $\eta/s$, which is calculated in the $\sigma$ model.  The main points are (i) that $C_{\sigma}$ is significantly larger than $C_{\pi}$, and (ii) they both depend on momentum.  They are certainly not constants which are equal to each other. Despite the fact that  $C_{\sigma} > C_{\pi}$ the shear viscosity is almost entirely determined by the pion because the $\sigma$ contribution is exponentially suppressed by the Boltzman factor.  It is easy to imagine from Fig. 2 that integrating $2 T^3 C_{\pi}$ with the weighting in Eq. (\ref{gasshear}) would give essentially the same result as integrating the momentum-independent $\eta/s$.  

\begin{figure*}[ht]
\centering
\includegraphics[width=0.48\textwidth]{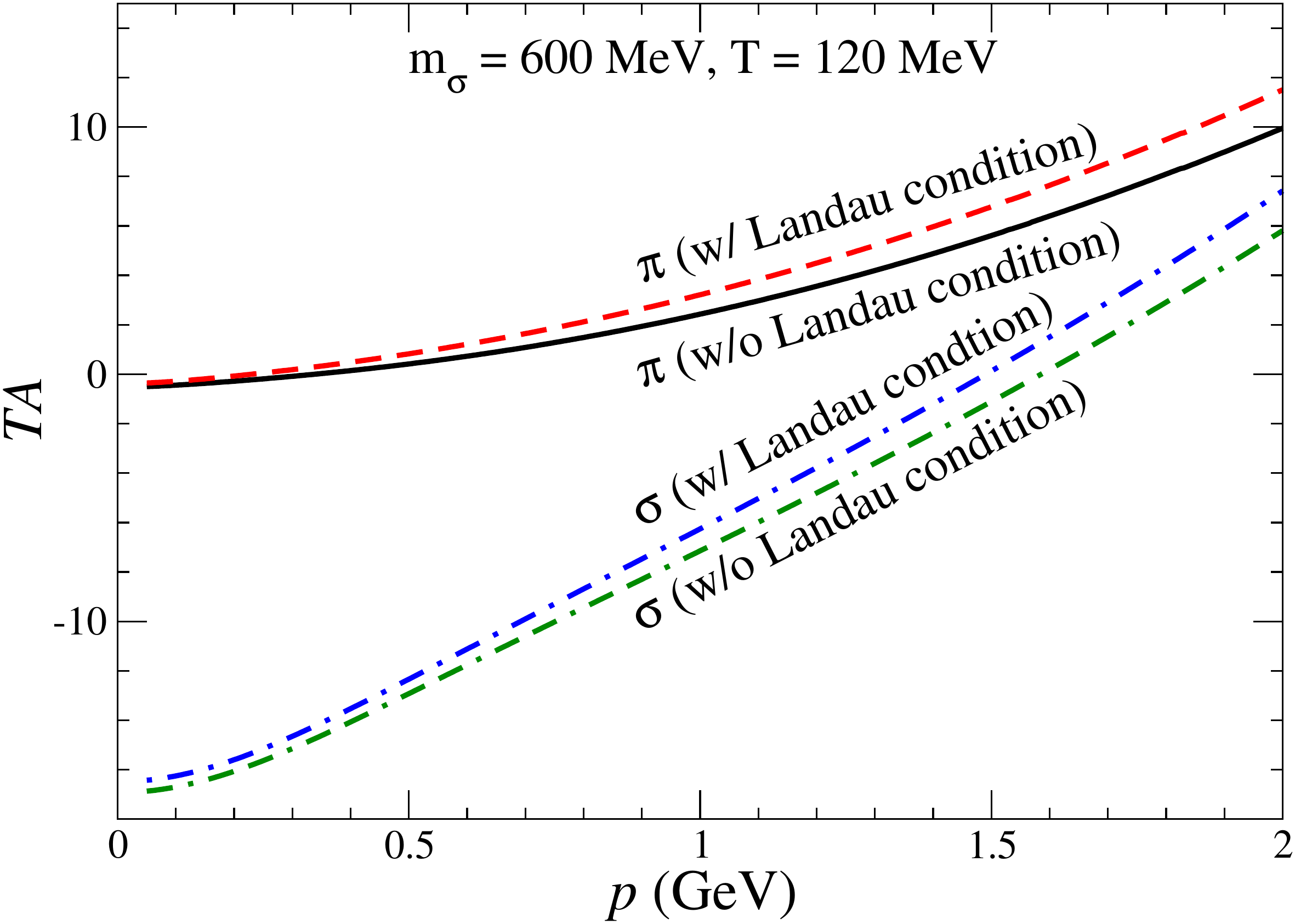}
\includegraphics[width=0.48\textwidth]{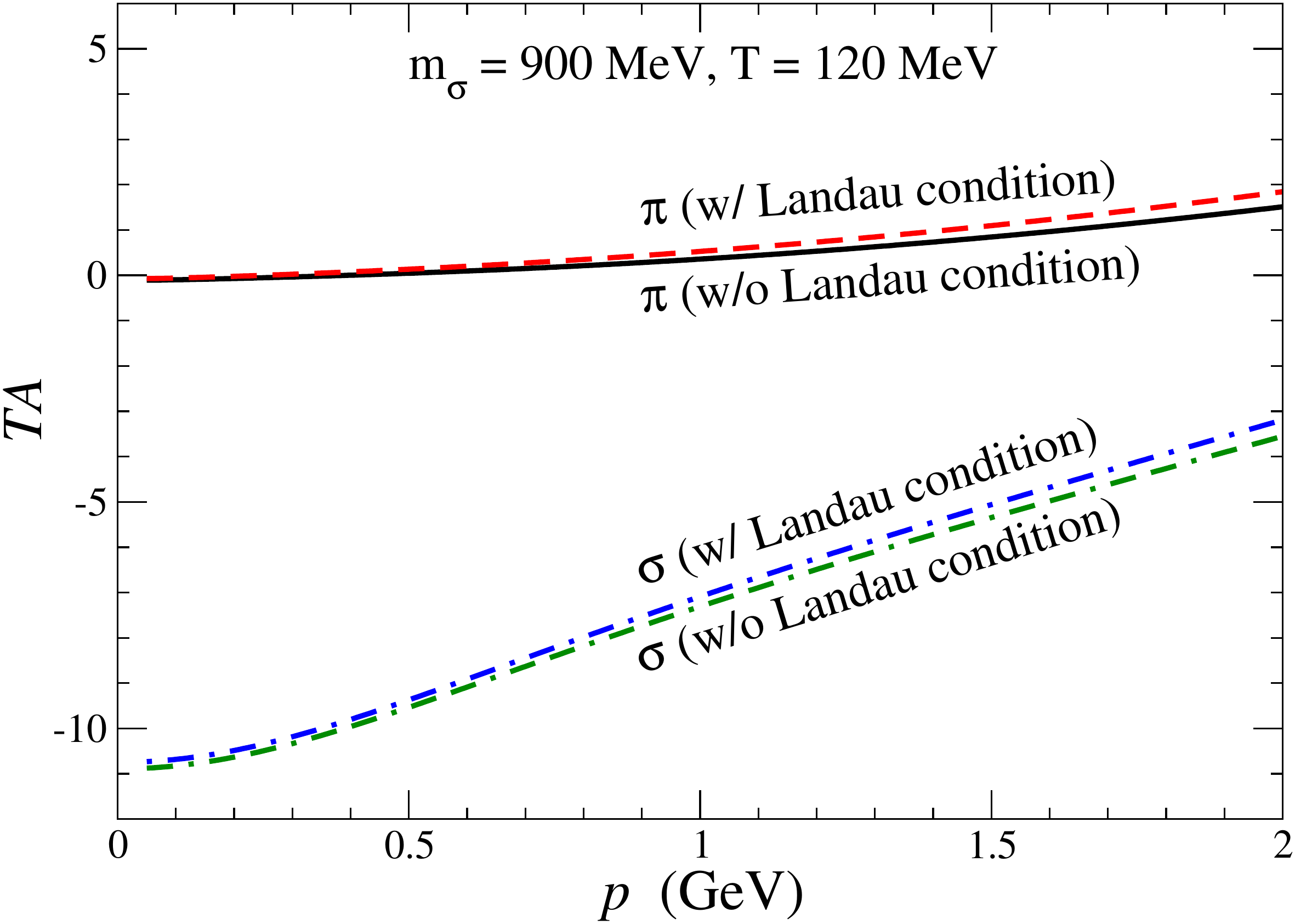}
\includegraphics[width=0.48\textwidth]{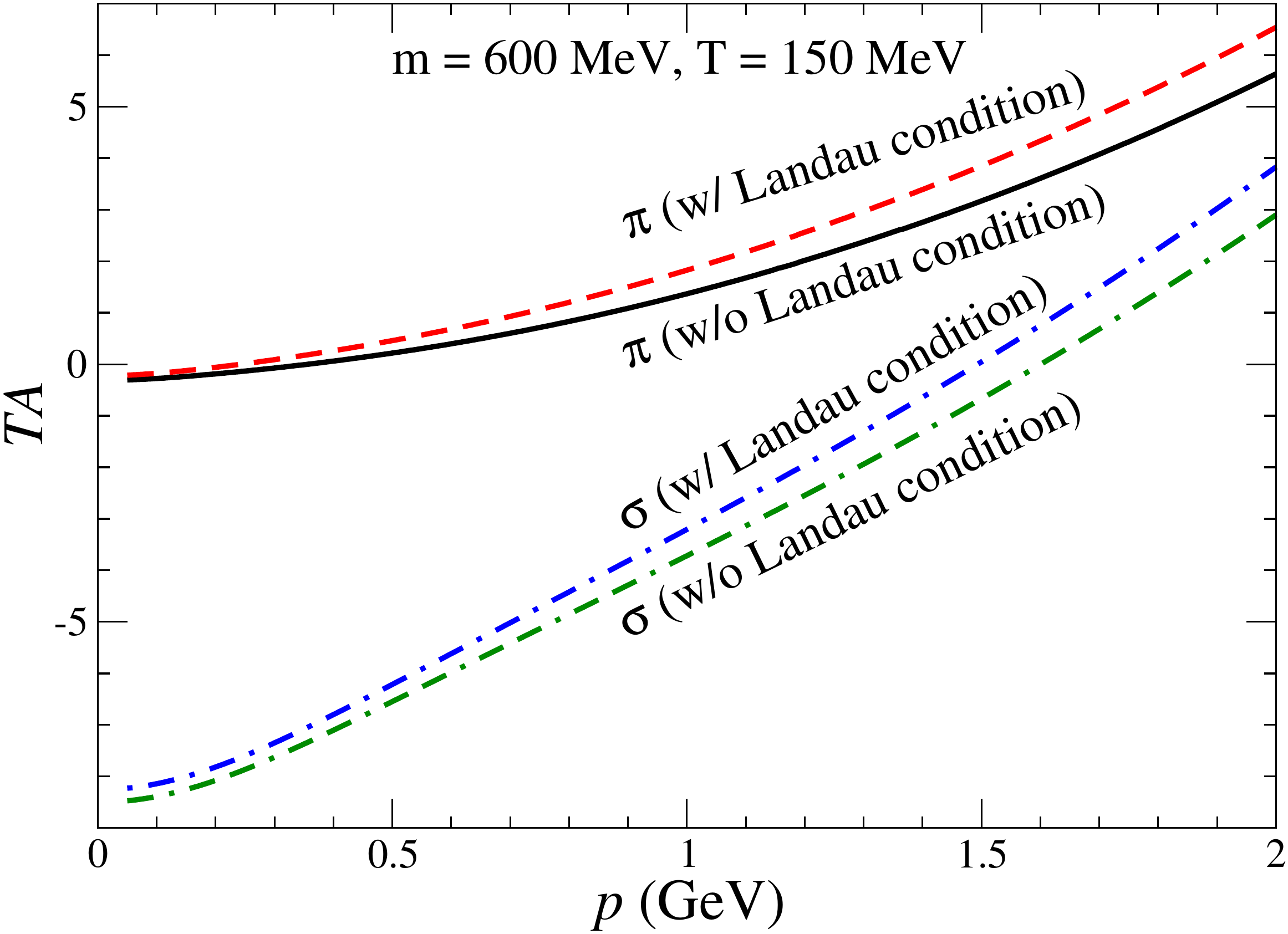}
\includegraphics[width=0.48\textwidth]{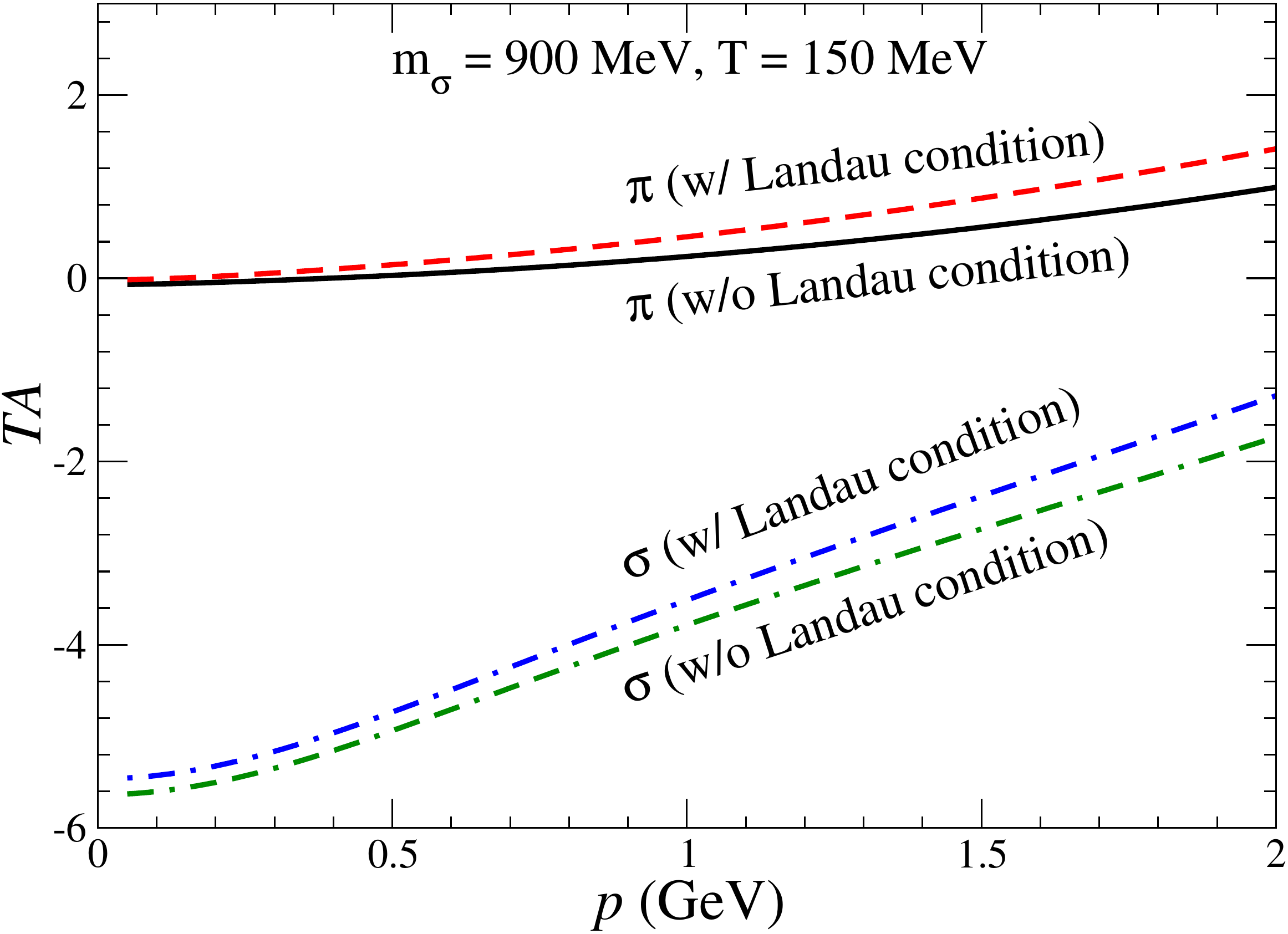}
\caption{The bulk functions $A_{\pi}$ and $A_{\sigma}$ in the linear $\sigma$ model for the indicated values of the vacuum $\sigma$ mass and temperature.}
\label{Fig3}
\end{figure*} 

\section{Bulk Viscosity}

The expression for the bulk viscosity is
\be
\zeta = \frac{1}{3} \sum_a \int \frac{d^3p}{(2\pi)^3} \frac{|{\bf p}|^2}{E_a} 
f_a^{\rm eq}(E_a/T) A_a(E_a) \,.
\label{gasbulk}
\ee
The Chapman-Enskog equation, which is an integral equation that determines the $A_a$ in quasi-particle models, does not have a unique solution.  Starting 
with one set of solutions $A_a(E_a)$ we can generate an infinite number of other solutions by making the shift $A_a(E_a) \rightarrow A'_a(E_a) = A_a(E_a) - bE_a$, where $b$ 
is an arbitrary constant.  This constant is associated with energy conservation.  Among the infinite number of solutions $A_a(E_a)$ one must choose that one which satisfies the Landa-Lifshitz condition, which means that $\boldsymbol{v}$ represents the velocity of energy flow.  (The Eckart condition means that $\boldsymbol{v}$ represents the velocity of baryon flow, which is obviously undefined if the system has zero net baryon number.)  This requirement is sometimes called the condition of fit.  If one has a particular solution $A^{\rm par}_a(E_a)$ then the condition of fit is
\be
\sum_a \int \frac{d^3p}{(2\pi)^3E_a} f_a^{\rm eq} 
\left[ E_a^2 - T^2 \frac{d \bar{m}_a^2}{dT^2} \right]
\left[ A^{\rm par}_a(E_a) - bE_a \right] = 0 \, .
\ee
With some manipulation the constant $b$ can be shown to be
\be
b = \frac{v_s^2}{T^2 s}
\sum_a \int \frac{d^3p}{(2\pi)^3E_a} f_a^{\rm eq} 
\left[ E_a^2 - T^2 \frac{d \bar{m}_a^2}{dT^2} \right] A^{\rm par}_a(E_a) \,,
\ee
where $v_s$ is the speed of sound.  If $A^{\rm par}_a(E_a)$ satisfies the condition of fit then $b=0$.  In general though
\ba
\zeta &=& \frac{1}{3} \sum_a \int \frac{d^3p}{(2\pi)^3 E_a} f_a^{\rm eq} A^{\rm par}_a(E_a) \nonumber \\
&\times& \left[ |{\bf p}|^2 - 3 v_s^2 \left( E_a^2-T^2 \frac{d\bar{m}_a^2}{dT^2} \right) \right] \,.
\label{bulkMFb} 
\ea
A particular, approximate, solution in the relaxation time approximation is
\be
A^{\rm par}_a(E_a) = \frac{\tau_a(E_a)}{3TE_a} \left[  |{\bf p}|^2 - 3 v_s^2 \left( E_a^2-T^2 \frac{d\bar{m}_a^2}{dT^2} \right) \right] \,.
\label{ArelaxMF}
\ee 
Substitution of Eq. (11) into Eq. (10) guarantees that the bulk viscosity is positive; see also \cite{Gavin:1985ph,JeonYaffe}.  In the relaxation time approximation the function that should appear in Eq. (2) is the one which satisfies the condition of fit
\be
A_a(E_a) = \frac{\tau_a(E_a)}{3TE_a} \left[  |{\bf p}|^2 - 3 v_s^2 \left( E_a^2-T^2 \frac{d\bar{m}_a^2}{dT^2} \right) \right] - b E_a \,.
\ee

In the first approximation one expects that the bulk viscosity is very small except near a phase transition or rapid crossover \cite{Kapusta:2008vb}.  Thus the default $A_a = 0$ as mentioned earlier.

In the next approximation one might consider all $A_a$ to be equal and independent of momentum, just as in the shear case.  Substitution into Eq. (7) yields
\be
A = \frac{\zeta}{P_0} \,.
\ee
However, this does not satisfy the condition of fit.  Substitution into Eq. (10) leads instead to
\be
A = \frac{\zeta}{P_0 - v_s^2 (\epsilon_0 + T dV/dT)} \,.
\ee
The denominator is in general very close to zero, and could be positive or negative or even vanish at some temperature.  For a nonrelativistic gas with $V=0$ it is negative.  To our knowledge this approximation has not been used in the literature.

Reference \cite{em-df2} assumed that all particles had the same momentum-independent relaxation time $\tau_R$.  Let us also assume that the masses are temperature-independent.  Then it is easily shown that $b = 0$ so that
\be
A_a(E_a) = \frac{\tau_R}{3TE_a} \left[  |{\bf p}|^2 - 3 v_s^2 E_a^2 \right] \,.
\ee
This expression is on the right track since the condition of fit is satisfied, but it is generally invalid ($b$ is not zero) when either of the assumptions are not met.  

Figure 3 shows $A_{\pi}$ and $A_{\sigma}$ as functions of momentum in the linear $\sigma$ model for the same combinations of $m_{\sigma}$ and $T$ as in previous figures.  The dependence on momentum is quite strong.  Enforcing the condition of fit ($b \neq 0$) results in only a minor modification of the numerical results.  This feature may not hold in more realistic models when a single species, in this case the pion, does not dominate the behavior of the viscosities.

\section{Conclusion}

In this paper we studied the applicability of the oft-used approximation embodied in Eq. (3) to high energy heavy ion collisions.  For illustrative purposes we used the linear $\sigma$ model with only pions and $\sigma$ mesons.  We showed how the shear functions  $C_a$ depend on momentum, and that their magnitudes vary from one particle species to another by factors of 2 or 3.  We also showed that the bulk functions $A_a$ have a strong dependence on momentum, and differ even more widely among particle species than the shear functions.  Use of Eq. (3) along with $A_a = 0$ is a rough first approximation to the physics, but for precision studies one must go beyond that.  For precision one should calculate the shear and bulk functions using the same particle species and same interactions which are used in the equation of state at the time of conversion from fluid to hadrons, and which are coded in the hadronic cascade (if an afterburner is used).  

In future work we intend to compare results obtained from the Chapman-Enskog integral equations with those in the relaxation time approximation.  A comparison of those approaches was performed in \cite{Wiranata1} for elastic $\pi$-$\pi$ collisions using various forms of the cross section, but only the resulting shear viscosities were calculated and the departure functions were not presented.  In \cite{Wiranata2} a resonance gas was considered with $K$-matrix cross sections, but again the focus was on the shear viscosity without presenting the corresponding departure functions.  Reference \cite{Jaiswal} calculated the departure function of pions to second order in gradients of the flow velocity with an energy-independent relaxation time determined by the ratio $\eta/s$.  Of course, one should include the usual list of all hadrons and resonances, with or without the inclusion of mean fields.  (The papers cited above did not include mean fields nor did they include the effects of bulk viscosity.)  Finally, nonzero chemical potentials can readily be accommodated as in \cite{ChemE}.

\section*{Acknowledgments}

We thank C. Gale and C. Shen for discussions.  This work was supported by the US Department of Energy (DOE) under Grant DE-FG02-87ER40328.


\begin{thebibliography}{39}

\bibitem{QM2014}
 Proceedings of Quark Matter 2014, Nucl. Phys. {\bf A931}, (2014).

\bibitem{QM2015}
 Proceedings of Quark Matter 2015, Nucl. Phys. {\bf A956}, (2016).

\bibitem{Teaney-df}
``The Effect of Shear Viscosity on Spectra, Elliptic Flow, and HBT Radii", D. Teaney, Phys. Rev. C {\bf 68}, 034913 (2003).

\bibitem{em-df1}
``Electromagnetic Radiation as a Probe of the Initial State and of Viscous Dynamics in Relativistic Nuclear Collisions", G. Vujanovic, J.-F. Paquet, G. S. Denicol, M. Luzum, S. Jeon, and C. Gale, Phys. Rev. C {\bf 94}, 014904 (2016). 

\bibitem{degroot}
S. R. de Groot, W. A. van Leeuwen and Ch. G. van Weert, {\it Relativistic Kinetic Theory: Principles and Applications} (North-Holland, Amsterdam, 1980).

\bibitem{MolnarWolff}
``Self-Consistent Conversion of a Viscous Fluid to Particles", D. Molnar and Z. Wolff, arXiv:1404.7850 (unpublished).

\bibitem{em-df2}
``The Production of Photons in Relativistic Heavy-Ion Collisions", J.-F. Paquet, C. Shen, B. Schenke, G. S. Denicol, M. Luzum, S. Jeon, and C. Gale, Phys. Rev. C {\bf 93}, 044906 (2016).

\bibitem{transport1}
``Theory of Shear and Bulk Viscosities of Hadronic Matter", P. Chakraborty and J. I. Kapusta, Phys. Rev. C {\bf 83}, 014906 (2011).

\bibitem{Gavin:1985ph} 
``Transport Coefficients In Ultrarelativistic Heavy Ion Collisions,'' S. Gavin, Nucl. Phys. A {\bf 435}, 826 (1985).

\bibitem{JeonYaffe}
``From Quantum Field Theory to Hydrodynamics: Transport Coefficients and Effective Kinetic Theory", S. Jeon and L. G. Yaffe, Phys. Rev. D {\bf 53}, 5799 (1996).

\bibitem{Kapusta:2008vb}
``Viscous Properties of Strongly Interacting Matter at High Temperature,''  J. I. Kapusta, in {\it Relativistic Heavy Ion Physics}, Landolt-B\"ornstein Group 1, Vol. 23 (Springer, Berlin, 2010).

\bibitem{Wiranata1}
``Shear Viscosities from the Chapman-Enskog and the Relaxation Time Approaches'', A. Wiranata and M. Prakash, Phys. Rev. C {\bf 85}, 054908 (2012).

\bibitem{Wiranata2}
``Shear Viscosity of Hadrons with K-matrix Cross Sections", A. Wiranata, V. Koch, M. Prakash and X. N. Wang,  Phys. Rev. C {\bf 88}, 044917 (2013).

\bibitem{Jaiswal}
``Relativistic Viscous Hydrodynamics for Heavy-Ion Collisions: A Comparison between the Chapman-Enskog and Grad Methods'', R. S. Bhalerao, A. Jaiswal, S. Pal, and V. Sreekanth, Phys. Rev. C {\bf 89}, 054903 (2014).

\bibitem{ChemE}
``Quasiparticle Theory of Transport Coefficients for Hadronic Matter at Finite Temperature and Baryon Density", M. Albright and J. I. Kapusta, Phys. Rev. C {\bf 93}, 014903 (2016).

\end{thebibliography}
\end{document}